\documentclass[10pt]{article}

\usepackage[margin=1in]{geometry}
\usepackage{booktabs}
\usepackage{graphicx}
\usepackage{xcolor}
\usepackage[numbers,sort&compress]{natbib}
\usepackage[hidelinks]{hyperref}
\usepackage{microtype}
\usepackage{tabularx}
\usepackage{authblk}
\usepackage{titlesec}
\usepackage{abstract}

\titleformat{\section}{\large\bfseries}{\thesection}{1em}{}
\titleformat{\subsection}{\normalsize\bfseries}{\thesubsection}{1em}{}

\title{\textbf{The Web-CLI: Verifiable Privacy for Tools, Models, and Inference Engines in the Browser}}

\author[]{Tejaswi Gowda}
\affil[]{Arizona State University, Tempe, AZ, USA. \texttt{tejaswi@asu.edu}}

 \date{}

\begin{document}

\maketitle

\begin{abstract}
We introduce the \emph{Web-CLI}, a novel application architecture deploying powerful computational capabilities (command-line tools compiled to WebAssembly, models run through client-side inference runtimes, and GPU-accelerated engines) as zero-install, offline-capable browser applications that preserve full underlying capability. Unlike traditional web-based alternatives that require server-side processing and expose user data to third parties, Web-CLI applications execute entirely on the client, providing a verifiable privacy guarantee by architectural necessity rather than policy. We define the Web-CLI pattern and enumerate its four design properties: \emph{fidelity}, \emph{progressive disclosure}, \emph{offline-first}, and \emph{zero egress}. We present four reference implementations spanning distinct computational domains: \texttt{ffmpeg-webCLI}, a full-featured browser-based video editor built on FFmpeg; \texttt{whisper-webCLI}, a Whisper-based speech transcription tool running via Transformers.js; \texttt{chat-webCLI}, a WebLLM-based language model inference interface; and \texttt{3mf-webCLI}, a deterministic browser tool that segments 3D models into multi-material files for physical 3D printing. Together these implementations demonstrate that the Web-CLI pattern generalizes across deterministic media processing, neural speech recognition, large language model inference, and geometry processing with a physical output, and we outline how the pattern extends to AI-native interfaces in which a local language model becomes the command surface itself. We further report early, anecdotal signs that the pattern has been independently reused by third-party tools, suggesting it generalizes beyond its reference implementations. We evaluate the primary reference implementation against native FFmpeg on performance and feature parity, and argue that the progressive-disclosure design lowers the barrier for non-technical users. We argue that for applications processing sensitive user data (medical, legal, journalistic, or personal), the Web-CLI should be considered the default architecture, as it makes data locality an independently verifiable technical property rather than a policy promise.
\end{abstract}


\section{Introduction}
%% ============================================================

The command line is one of the most powerful interfaces ever designed. Tools like FFmpeg, ImageMagick, Pandoc, and Sox encode decades of engineering expertise into a single executable capable of transforming, analyzing, and processing complex media with precision that graphical applications rarely match. Yet this power comes with a steep price: fluency in the terminal, knowledge of flags and filter syntax, and the ability to install and configure software on a local machine. For the vast majority of users, this price is too high.

The web has long promised to close this gap. Browser-based tools for video editing, image processing, and document conversion have proliferated, but at a cost rarely made explicit. Every file uploaded to a cloud-based video editor leaves the user's device and is processed on a remote server owned by a third party. For most use cases this is acceptable. For a significant and growing class of use cases (medical video, legal evidence, journalistic source material, personal and intimate media), it is not. The privacy guarantee offered by these services is a policy document, not a technical property. It can be changed, violated, subpoenaed, or breached.

WebAssembly changes the calculus. Introduced in 2017~\cite{haas2017bringing} and standardized as a W3C recommendation in 2019~\cite{wasmspec}, WebAssembly has matured into a viable compilation target for complex, performance-sensitive C and C++ codebases. Projects like \texttt{ffmpeg.wasm}~\cite{ffmpegwasm} have demonstrated that tools previously considered too computationally intensive for the browser can execute entirely on the client at practical speeds. This opens a possibility that has not yet been fully articulated as an architectural pattern: the deployment of CLI tools as browser applications that require no installation, operate offline, and process user data exclusively on the user's own hardware.

We call this the \emph{Web-CLI} architecture. A Web-CLI application is defined by four properties: (1)~\emph{fidelity}: it exposes the full capability of the underlying CLI tool without artificial restriction; (2)~\emph{progressive disclosure}: it provides a graphical interface for common operations while preserving access to the raw command interface for power users; (3)~\emph{offline-first}: after an initial load it requires no network connection to function; and (4)~\emph{zero egress}: user data never leaves the device, a guarantee enforced by the architecture rather than by policy.

These four properties together constitute something qualitatively different from existing browser-based tools. A Web-CLI application is not a simplified cloud wrapper around a powerful tool; it \emph{is} the powerful tool, running locally, wrapped in an interface that makes it accessible to users who could never have used it before.

We present \texttt{ffmpeg-webCLI}\footnote{\url{https://tejaswigowda.com/ffmpeg-webCLI/}}~\cite{ffmpegwebcli} as the primary reference implementation of the Web-CLI architecture. Built on the \texttt{ffmpeg.wasm} WebAssembly port of FFmpeg~\cite{ffmpegwasm}, it exposes 30+ video processing operations through a graphical interface (trim, compress, convert, GIF creation, audio extraction, subtitle embedding, picture-in-picture, side-by-side composition, logo overlay, and more) while simultaneously offering a raw FFmpeg command mode in which users can type any valid FFmpeg arguments and receive a live preview of the exact command that will execute. An embedded example command library bridges the two modes, offering one-click recipes for operations like loudness normalization, video stabilization, and lossless remuxing.

The tool requires no installation, no account creation, and no network access after the initial 31\,MB WebAssembly binary is cached by the browser. It runs on any modern browser across any operating system. It works on a plane. It works in a hospital. It works in an air-gapped environment. And it works for a non-technical user who has never opened a terminal.

We supplement this primary case study with three more reference implementations: \texttt{whisper-webCLI}\footnote{\url{https://tejaswigowda.com/whisper-webCLI/}}~\cite{whisperwebcli} for speech recognition (running Whisper via Transformers.js~\cite{transformersjs}), \texttt{chat-webCLI}\footnote{\url{https://tejaswigowda.com/chat-webCLI/}}~\cite{chatwebcli}, a local chat application for language model inference (on WebLLM~\cite{webllm}), and \texttt{3mf-webCLI}\footnote{\url{https://tejaswigowda.com/3mf-webCLI/}}~\cite{tdmfwebcli}, a deterministic tool that prepares 3D models for multi-material physical printing. These demonstrate that the Web-CLI pattern scales from deterministic media processing through neural inference and large language model deployment to geometry processing with a physical output, and sketch how it further extends to AI-native interfaces in which the local model itself becomes the command surface. We have additionally observed early, anecdotal signs of independent reuse by third-party tools.

The contributions of this paper are as follows:

\begin{itemize}
  \item We define the \emph{Web-CLI} architecture and enumerate its four defining properties: fidelity, progressive disclosure, offline-first, and zero egress.
  \item We identify the class of tools, models, and inference engines amenable to this architecture and propose criteria for evaluating suitability.
  \item We present \texttt{ffmpeg-webCLI} as a reference implementation, describing its design decisions, feature set, and the engineering challenges of full-fidelity Web-CLI deployment.
  \item We demonstrate the pattern's generalizability across four reference implementations spanning execution substrates and output modalities: deterministic media processing (FFmpeg on WASM), speech recognition (Whisper via Transformers.js/ONNX), language model inference (WebLLM on WebGPU), and geometry processing with a physical output (\texttt{3mf-webCLI}, GLB-to-3MF multi-material segmentation for 3D printing). We show that these compose into multi-stage on-device pipelines without breaking the privacy guarantee (local auto-captioning), identify AI-native Web-CLIs as the pattern's principal future direction, and report early anecdotal signs of independent reuse by third parties.
  \item We evaluate the primary reference implementation against native FFmpeg on performance and feature parity, and argue that progressive disclosure lowers the barrier for non-technical users.
  \item We argue that for applications processing sensitive user data, the Web-CLI should be considered the default architecture; not because it is always the most performant option, but because it is the only option that makes privacy a technical guarantee rather than a policy promise.
\end{itemize}

The remainder of this paper is organized as follows. Section~\ref{sec:related} surveys related work. Section~\ref{sec:architecture} defines the Web-CLI architecture. Section~\ref{sec:implementation} describes the \texttt{ffmpeg-webCLI} implementation. Section~\ref{sec:casestudies} presents the speech-recognition, language-model, and 3D-printing case studies. Section~\ref{sec:evaluation} presents our evaluation. Section~\ref{sec:discussion} discusses generalizability and limitations. Section~\ref{sec:futurework} describes future work. Section~\ref{sec:conclusion} concludes.

%% ============================================================
\section{Related Work}
\label{sec:related}
%% ============================================================

\subsection{WebAssembly as a Compilation Target}

WebAssembly (WASM) was introduced in 2017 as a portable, low-level bytecode format for the web~\cite{haas2017bringing} and reached W3C Recommendation status in 2019~\cite{wasmspec}. It was designed as a compilation target for languages like C, C++, and Rust, enabling near-native execution speeds in the browser. Emscripten~\cite{emscripten} is the primary toolchain for compiling C/C++ codebases to WASM, and has been used to port a wide range of systems software to the browser.

Prior work has rigorously analyzed WASM's performance for computationally intensive workloads relative to native code~\cite{jangda2019} and demonstrated its viability for production systems software such as database engines~\cite{sqlitewasm}. SQLite's official WASM port~\cite{sqlitewasm} is a notable prior instantiation of what we call the Web-CLI pattern, demonstrating that production-grade systems software can be deployed in the browser without meaningful loss of capability.

\subsection{Browser-Based Media Processing}

Existing browser-based video editors (including Kapwing, Clideo, and Cloudconvert) process user files on remote servers. This architecture introduces latency proportional to upload speed, requires ongoing server infrastructure, and creates privacy risks for users handling sensitive content. Academic treatments of web-based media have largely concerned delivery and streaming protocols~\cite{streamingweb} rather than privacy-preserving client-side processing architectures.

\texttt{ffmpeg.wasm}~\cite{ffmpegwasm} established the feasibility of running FFmpeg in the browser but did not address the user interface layer or articulate an architectural pattern. Our work builds on this foundation by contributing the Web-CLI abstraction and a full reference implementation.

\subsection{Privacy-Preserving Computation}

Privacy-preserving computation has been studied extensively in the context of cryptographic protocols: secure multi-party computation~\cite{smpc}, homomorphic encryption~\cite{he}, and differential privacy~\cite{dp}. These approaches are powerful but impose significant computational overhead and are ill-suited to media processing workloads.

Client-side computation as a privacy mechanism has received less formal attention. Work on local-first software~\cite{localfirst} argues for architectures in which user data resides on the user's own device, but does not specifically address the deployment of existing CLI tools via WASM. The Web-CLI architecture can be understood as a specific instantiation of the local-first principle for the domain of CLI tools.

\subsection{Democratization of Technical Tools}

A substantial body of HCI research on end-user programming and user-interface design has examined how interfaces broaden access to computational tools for non-technical users~\cite{enduser1, enduser2}. GUI wrappers around CLI tools (including Audacity, Handbrake, and ImageOptim) have demonstrated that interface design can dramatically expand the user population of a tool without reducing its power. The Web-CLI extends this tradition by eliminating the installation barrier entirely.

%% ============================================================
\section{The Web-CLI Architecture}
\label{sec:architecture}
%% ============================================================

\subsection{Definition}

We define a \emph{Web-CLI application} as a browser-based application that deploys a powerful computational capability (a command-line tool compiled to WebAssembly, a model executed through a client-side inference runtime, or a GPU-accelerated engine) entirely on the user's device, satisfying four properties:

\begin{enumerate}
  \item \textbf{Fidelity.} The application exposes the full computational capability of the underlying engine. No capability available in the native tool or model is artificially restricted in the browser deployment. In particular, a Web-CLI application must provide a \emph{raw control surface}: for a tool-based engine this is a command passthrough in which users invoke the tool with arbitrary arguments; for a model-based engine it is direct access to the raw inference parameters (model selection, language, decoding settings, and so on).

  \item \textbf{Progressive Disclosure.} The application provides a graphical interface for common operations, enabling non-technical users to accomplish tasks without knowledge of the underlying syntax or parameters. This interface coexists with, rather than replaces, the raw control surface, allowing power users to access the full capability of the engine.

  \item \textbf{Offline-First.} After an initial load (including download of the WASM binary, inference runtime, and any required model weights or assets), the application functions without network connectivity. No operation requires a server round-trip. This property is relaxed to \emph{zero-egress-after-load} for applications (such as large language model inference) where model weights are too large for practical pre-bundling but are cached after first download.

  \item \textbf{Zero Egress.} User data (input files, intermediate outputs, prompts, audio, and final results) never leaves the user's device. This guarantee is enforced by the architecture: the application has no server endpoint to which user data could be transmitted; only the engine, runtime, and model assets are fetched, and only once. It is a technical property, not a policy commitment, and can be verified empirically by inspecting network requests during operation.
\end{enumerate}

Two observations follow from this definition. First, the architecture is \emph{substrate-agnostic}: the four properties make no reference to how the capability is realized, and our reference implementations deliberately span multiple execution substrates (a CLI binary compiled to WASM, a neural model run through a JavaScript/ONNX inference runtime, an LLM executed through a WebGPU runtime, and a deterministic JavaScript geometry pipeline) across as many domains. Second, the ``CLI'' in Web-CLI denotes the \emph{raw control surface} rather than a literal command line: it is a textual command interface for tool-based engines and a raw-parameter interface for model-based engines. What unifies the family is not a shared toolchain but the combination of full on-device capability, a layered interface from preset to raw control, and an architecturally-enforced zero-egress guarantee.

\paragraph{On the name.} We retain the term ``Web-CLI'' deliberately, though we use ``CLI'' in a generalized sense. Historically the command line denotes a mode of interaction: direct, full-fidelity, expert-level control of a tool through a textual interface, rather than merely a particular shell. It is this connotation we invoke: the ``CLI'' of a Web-CLI is the \emph{raw control surface} that progressive disclosure exposes beneath the graphical layer, giving expert users unmediated access to the underlying capability. For a tool-based engine such as FFmpeg, that surface is literally a command line of arguments. For a model-based engine such as Whisper or an LLM, it is the raw inference parameters (model, language, decoding settings, system prompt). In both cases the role is identical: the expert-access tier beneath the GUI, even though its concrete form differs. We therefore treat ``CLI'' as a metaphor for this raw-control tier, not as a claim that every Web-CLI wraps a literal command-line program.

\subsection{Architecture Overview}

Figure~\ref{fig:architecture} illustrates the Web-CLI architecture. The core components are:

\begin{figure}[h]
  \centering
  \includegraphics[width=0.95\linewidth]{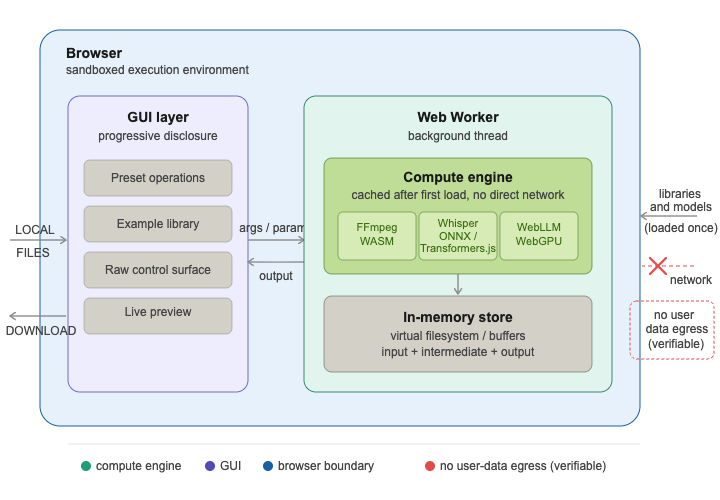}
  \caption{The substrate-agnostic Web-CLI architecture. A progressive-disclosure GUI drives a compute engine (a WASM binary, an inference runtime, or a WebGPU runtime) running in a Web Worker, with access only to an in-memory store. Local files enter and outputs download on-device; engine and model assets are fetched once on first load. User data does not egress: a property that is independently verifiable rather than physically impossible.}
  \label{fig:architecture}
\end{figure}

\textbf{Compute Engine.} The capability is realized by a compute engine that executes on the client: a CLI tool compiled to WebAssembly via Emscripten (as with FFmpeg), a model run through a client-side inference runtime (as with Whisper via Transformers.js), or a GPU-accelerated runtime (as with WebLLM on WebGPU). The engine and its assets are loaded once and cached by the browser. It executes in a sandboxed environment with no network access and no filesystem access beyond the in-memory virtual filesystem provided by the runtime.

\textbf{Web Worker.} The compute engine executes in a Web Worker, a browser thread separate from the main UI thread. This prevents blocking the UI during long-running operations and enables progress reporting via message passing.

\textbf{Virtual Filesystem.} Input data is written to an in-memory virtual filesystem (or equivalent in-memory buffer) before being passed to the engine. Outputs are read from the same store after processing. No data touches the real filesystem or any network endpoint.

\textbf{GUI Layer.} A web-native interface provides controls for common operations. Parameters are mapped to engine arguments or inference settings, which are assembled and passed to the engine. For tool-based engines a live command preview shows the exact invocation that will execute, bridging the GUI and raw modes.

\textbf{Raw Control Surface.} A raw interface allows users to drive the engine directly: arbitrary command arguments for tool-based engines, or direct inference parameters for model-based engines. The same execution path is used as for GUI-generated operations, ensuring complete fidelity.

\textbf{Example Library.} A curated set of templates for common but complex operations. Each template pre-fills the raw interface with a working invocation, serving as a learning resource as well as a productivity tool.

\subsection{Suitability Criteria}

Not all CLI tools are equally amenable to the Web-CLI architecture. We propose the following criteria for assessing suitability:

\textbf{WASM compilability.} The tool must be compilable to WebAssembly. Tools with dependencies on OS-specific APIs (e.g., raw socket access, kernel interfaces) or GPU hardware may require modification. Most C/C++ codebases that avoid platform-specific syscalls are compilable with moderate effort.

\textbf{Memory footprint.} WASM applications are constrained by browser memory limits (typically 2--4\,GB). Tools processing large files may hit these limits. FFmpeg handles this gracefully by processing in streaming fashion; tools that require the entire input to be in memory simultaneously are less suitable.

\textbf{Binary size.} The WASM binary must be downloaded before the application can function. Binaries under $\sim$50\,MB are practical; larger binaries require careful loading strategies (lazy loading, streaming compilation) and clear user communication about initial load time.

\textbf{Computational intensity.} WASM executes meaningfully slower than native for CPU-bound workloads. The per-instruction penalty is modest: prior work reports roughly $1.5$--$2.5\times$~\cite{jangda2019}, and we measure $1.3$--$1.4\times$ on operations that are single-threaded natively. The gap compounds, however, when the WASM build cannot use the parallelism native tools enjoy: our single-threaded H.264 re-encodes run $25$--$33\times$ slower than multithreaded software x264 and $72$--$90\times$ slower than hardware-accelerated native encoding (Section~\ref{sec:evaluation}). Operations taking seconds natively may take minutes in WASM; suitability depends on operation type and clip length.

\textbf{Stateless operation.} CLI tools that operate statelessly (read inputs, produce outputs, exit) map cleanly to the Web-CLI pattern. Tools requiring persistent daemon processes or inter-process communication are significantly harder to port.

\subsection{Privacy Guarantees}

The zero-egress property of the Web-CLI architecture provides a privacy guarantee that is qualitatively different from policy-based guarantees offered by cloud services. Specifically:

\begin{itemize}
  \item \textbf{The compute engine has no network access.} The WASM module (or inference runtime) that processes user data executes in a sandbox with no direct network access; it cannot itself initiate a transmission. The surrounding application, like any web page, \emph{could} in principle open a network channel (Section~\ref{sec:platform}). Zero egress is therefore a property the application establishes by not transmitting user data, rather than a physical impossibility of the platform.
  \item \textbf{The guarantee is independently verifiable.} This is the crucial distinction from policy-based guarantees: any user can open browser developer tools and inspect the Network tab to confirm that no user data is transmitted during processing. The claim is not ``trust us''; it is directly checkable, in seconds, by anyone.
  \item \textbf{The guarantee survives service changes.} A cloud provider's privacy policy can change retroactively; a static, inspectable client-side application's behavior cannot change without a visible redeployment that users can re-verify.
  \item \textbf{Legal protection follows.} Data that never leaves a device cannot be subpoenaed from a third-party service provider. This is materially important for legal, medical, and journalistic use cases.
\end{itemize}

%% ============================================================
\section{Reference Implementation: ffmpeg-webCLI}
\label{sec:implementation}
%% ============================================================

\subsection{Overview}

\texttt{ffmpeg-webCLI} is a browser-based video editor implementing the Web-CLI architecture on top of FFmpeg via \texttt{ffmpeg.wasm}~\cite{ffmpegwasm}. It exposes 30+ video processing operations through a web-native graphical interface while preserving full access to FFmpeg's command-line interface. Beyond the preset operations exposed through dedicated GUI controls, it also provides an example command library (13 recipes), a raw command mode, and media-inspection utilities.

\subsection{Feature Set}

Table~\ref{tab:features} lists the operations available through the graphical interface, grouped by category.

\begin{table}[h]
\caption{Operations available in \texttt{ffmpeg-webCLI}}
\label{tab:features}
\begin{tabularx}{\linewidth}{lX}
\toprule
\textbf{Category} & \textbf{Operations} \\
\midrule
Conversion & Format convert (MP4, WebM, MKV, MOV, AVI), GIF maker, Audio extraction (MP3, AAC, WAV, OGG, FLAC) \\
\midrule
Editing & Trim, Crop, Rotate/Flip, Resize, Speed change (0.25$\times$--4$\times$), Reverse, Fade in/out \\
\midrule
Audio & Mute, Volume adjust (0--4$\times$), Audio replacement, Mix audio, Loop \\
\midrule
Enhancement & Brightness/Contrast/Saturation, Strip metadata \\
\midrule
Compositing & Logo overlay, Picture-in-picture, Side-by-side, Concatenate \\
\midrule
Subtitles & Embed subtitles (soft, MP4/MKV) \\
\midrule
Effects & Denoise (hqdn3d, three presets), Sharpen/Blur (unsharp/boxblur), Boomerang \\
\midrule
Social & Pad/Letterbox (16:9, 9:16, 1:1, 4:3, 4:5, 21:9), Normalize audio (EBU R128~\cite{ebu_r128}) \\
\midrule
Advanced & Raw FFmpeg command mode, Example command library (13 recipes), Media Info deep scan \\
\bottomrule
\end{tabularx}
\end{table}

\subsection{Progressive Disclosure Design}

A central design challenge in the Web-CLI architecture is presenting a tool of FFmpeg's complexity to users who may have no knowledge of its syntax. We address this through a three-level progressive disclosure model:

\textbf{Level 1: Preset operations.} Each of the 30+ preset operations in Table~\ref{tab:features} is exposed through a dedicated interface with appropriately constrained controls (sliders, dropdowns, file pickers). Users need no knowledge of FFmpeg syntax to use these operations.

\textbf{Level 2: Example command library.} A collapsible panel provides 13 curated FFmpeg command templates for operations not available through the preset interface, including loudness normalization (\texttt{loudnorm}), video stabilization (\texttt{deshake}), denoising (\texttt{hqdn3d}), and lossless remuxing (\texttt{-c copy}). Each template is pre-filled into the raw command interface with one click.

\textbf{Level 3: Raw command interface.} Users may type arbitrary FFmpeg arguments directly. A live command preview updates as they type, showing the exact command that will execute (including the \texttt{-i input} prefix and output filename). Quoted arguments containing spaces are handled correctly.

\subsection{Technical Implementation}

The application is implemented as a static web application with no server-side component. Key technical decisions:

\textbf{Web Worker isolation.} FFmpeg executes in a dedicated Web Worker, preventing UI blocking during processing. Progress is reported via \texttt{postMessage} callbacks.

\textbf{Execution core and isolation headers.} The deployed application uses the single-threaded \texttt{@ffmpeg/core} build (FFmpeg 5.1.4 compiled with Emscripten, SIMD enabled, pthreads disabled). \texttt{ffmpeg.wasm} also provides a multi-threaded core, which requires \texttt{SharedArrayBuffer} and therefore Cross-Origin Isolation (\texttt{COOP: same-origin} and \texttt{COEP: require-corp} headers). The application ships a minimal Node.js server (\texttt{server.js}) that sets these headers for local deployment, and the hosted version is served with them, so the multi-threaded core remains a drop-in upgrade. As Section~\ref{sec:evaluation} shows, single-threaded execution is the dominant cost for re-encoding operations.

\textbf{Trim composability.} The timeline trim range is applied as a preprocessing step to every operation that supports it, allowing users to combine trimming with any other operation (e.g., trim a clip then convert to GIF) without multiple processing passes.

\textbf{Live size estimation.} For compression operations, the UI provides a real-time file size estimate that updates as the user adjusts CRF and preset parameters. This is computed from the source file size and the expected compression ratio for the selected settings.

\textbf{Multi-input operations.} Several operations (audio replacement, mix audio, logo overlay, picture-in-picture, side-by-side, concatenate) require a second input file. These are handled by writing both files to the WASM virtual filesystem before invoking FFmpeg with appropriate \texttt{-i} arguments.

\textbf{Auto-captioning: a composed pipeline.} \texttt{ffmpeg-webCLI} imports the \texttt{transcriber.js} module from \texttt{whisper-webCLI} (Section~\ref{sec:casestudies}) to offer automatic captioning entirely on-device. The flow composes two Web-CLI engines: FFmpeg extracts the audio track to 16\,kHz mono; the Whisper engine transcribes it locally into timestamped segments, which are formatted as SRT and presented for editing; FFmpeg then muxes the (possibly edited) subtitles back into the video as a soft track. Because both stages run on the client and data passes between them through browser memory rather than any network, the composition preserves the zero-egress guarantee end to end. To remain within the WASM memory ceiling, the Whisper engine is disposed before FFmpeg re-engages for muxing. This demonstrates that Web-CLI applications compose into multi-stage local pipelines without sacrificing their privacy guarantees.

\textbf{Composition with native web APIs.} A second composition illustrates how Web-CLI tools draw on the broader web platform (Section~\ref{sec:platform}). The standard \texttt{ffmpeg.wasm} build omits \texttt{libass}, so subtitles cannot be hard-burned into the picture through FFmpeg's native subtitle filter. To support burned-in captions for platforms that require them (e.g., social video), the application renders each caption to a transparent image using the browser's Canvas API (with stroked text and a background band for legibility) and composites these timed images onto the video via FFmpeg's \texttt{overlay} filter.

\subsection{Progressive Web App and Offline-First Deployment}

\texttt{ffmpeg-webCLI} is deployed as a Progressive Web App (PWA), strengthening the offline-first property from passive to active. The distinction is significant: passive offline-first means the application happens to work offline if the user has previously visited; active offline-first means the application is explicitly engineered to work offline by design.

The PWA implementation comprises three components. First, a \textbf{service worker} intercepts all network requests and serves static assets (HTML, CSS, JavaScript) from cache, ensuring the application loads instantly without network access after the first visit. Second, the service worker proactively caches the \texttt{ffmpeg-core.wasm} binary (31\,MB) and all CDN dependencies on first load, so that subsequent offline use requires no network access whatsoever. Third, the Web app manifest enables \textbf{installation} on any platform (desktop or mobile), allowing users to add the application to their home screen or app drawer, where it launches as a standalone app without browser chrome.

A \textbf{Screen Wake Lock} API integration prevents device sleep during long encoding operations, which is particularly important on mobile devices where screen timeout may interrupt a multi-minute WASM encoding job.

The PWA deployment has a significant implication for the privacy argument. An installed PWA that operates offline has no ambient network context during processing: there is no active network interface over which data could be transmitted even in the presence of a compromised dependency. This represents a stronger privacy guarantee than a browser tab operating in an online context, where the browser's network stack remains active even if the application itself makes no requests.

The COOP/COEP header requirement for \texttt{SharedArrayBuffer} interacts non-trivially with service worker caching: cached responses served by the service worker must themselves include the correct isolation headers, which service workers do not automatically preserve. We address this by explicitly setting headers in the service worker's \texttt{fetch} handler for all cached responses, ensuring Cross-Origin Isolation is maintained in offline mode.

\subsection{Limitations}

The following limitations apply to the current implementation:

\textbf{Hard subtitle burning.} Burning subtitles into the video frame (as opposed to embedding them as a soft track) requires a \texttt{libass}-enabled FFmpeg build. The standard \texttt{ffmpeg.wasm} build does not include \texttt{libass}; this operation is therefore accomplished by rendering captions to images via the Canvas API and compositing them with FFmpeg. This is a workaround rather than a native solution, and it does not support advanced \texttt{libass} features like karaoke effects or complex positioning.

\textbf{Memory ceiling for processing.} Loading, previewing, and trimming setup work for files of any size, as inputs are referenced via object URLs rather than copied into memory. \emph{Processing}, however, remains bounded by the WASM runtime's practical memory limit of approximately 2\,GB: operations that require the entire decoded video in memory (notably whole-clip reversal), memory-hungry encoders and filters (1080p VP9 encoding fails at any length, and GIF conversion of a 10-minute clip exhausts memory in \texttt{palettegen}; Section~\ref{sec:evaluation}), or very large inputs may exhaust this budget and crash the tab, and are unsuitable for batch application across many files. This is a fundamental constraint of the current 32-bit WASM address space rather than a property of the application, and users are not yet warned of it proactively.

\textbf{Encoding performance.} With the deployed single-threaded core, H.264 re-encoding is $25$--$33\times$ slower than multithreaded software x264 and $72$--$90\times$ slower than hardware-accelerated native encoding on the same machine (Section~\ref{sec:evaluation}); stream-copy and audio operations run at or near native speed. Re-encoding short clips remains practical (a 60-second compress takes about five minutes); long re-encodes are currently impractical without a multi-threaded core or a GPU encoding path.

\textbf{No persistent history.} Processed files are not retained between sessions. Users must download output immediately after processing or reprocess from the original source.

\noindent\textit{Community validation.} A public release of \texttt{ffmpeg-webCLI} on Hacker News received 86 points and 43 comments within 24 hours~\cite{hn2026}, and the project accumulated over 1000+ GitHub stars and 100+ forks within weeks. Users reported successful real-world use including legacy format migration (.mpg to .mp4) and classroom deployment as an offline PWA. Community feedback drove active subsequent development: two of the most-requested capabilities (single-pass operation chaining and batch processing across multiple files) were implemented in response to public feature requests, and user-reported issues were diagnosed and resolved, including a UI gate that had prevented large files from loading (now handled via object-URL references, allowing multi-gigabyte inputs to load and preview without copying into memory).

%% ============================================================
\section{Case Studies}
\label{sec:casestudies}
%% ============================================================

\subsection{Whisper: Speech Recognition}

Whisper~\cite{whisper} is an open-source automatic speech recognition model developed by OpenAI. Several toolchains run Whisper in the browser: \texttt{whisper.cpp}~\cite{whispercpp}, a C/C++ port compiled to WebAssembly, and Transformers.js~\cite{transformersjs}, which runs Whisper as an ONNX model on a WebAssembly or WebGPU backend. Our implementation uses the latter.

We built \texttt{whisper-webCLI}~\cite{whisperwebcli}, a Web-CLI instantiation for speech recognition, as our second reference implementation. The transcription engine is implemented as a self-contained module (\texttt{transcriber.js}) running Whisper via Transformers.js; the same module is reused unmodified by \texttt{ffmpeg-webCLI} to provide automatic captioning (Section~\ref{sec:implementation}), demonstrating that Web-CLI \emph{components}, not only whole applications, compose across the family. It satisfies all four Web-CLI properties. Fidelity is preserved through access to model selection (tiny, base, small, medium), language specification across approximately 99 languages, optional translation to English, and output formats (plain text, SRT, VTT); the raw control surface here is the set of inference parameters (model, language, decoding temperature, beam size) rather than command-line arguments. Progressive disclosure is achieved through a simple drag-and-transcribe interface for common cases, with those advanced parameters and an editable transcript exposed for power users. Offline-first operation is achieved after the model weights and inference runtime are cached in the browser. This follows the same cached-after-download pattern as the FFmpeg binary. Zero egress is guaranteed by construction: audio never leaves the device, verifiable via the browser's network inspector.

The privacy implications of client-side speech recognition are particularly significant. Audio recordings of medical consultations, legal proceedings, therapy sessions, and journalistic interviews represent some of the most sensitive personal data in existence. Cloud-based transcription services require this data to be transmitted to and processed on remote servers. A Web-CLI transcriber eliminates this exposure entirely.

Whisper represents a different computational profile than FFmpeg: rather than deterministic filter-based processing, it performs neural network inference. That the very same module powers both a standalone transcriber and \texttt{ffmpeg-webCLI}'s auto-captioning shows that reuse in the Web-CLI family operates at the level of components, not only whole applications.

\subsection{Local LLM Inference}

WebLLM~\cite{webllm} is an open-source engine that enables large language model inference entirely in the browser via WebGPU acceleration. It supports a range of models including Llama, Mistral, Qwen, and Phi variants, and exposes an OpenAI-compatible API that runs entirely client-side.

We built a third reference implementation, \texttt{chat-webCLI}~\cite{chatwebcli}, a browser-based chat application that runs language model inference entirely on the client via WebLLM and WebGPU. It introduces an important refinement to the architecture. The offline-first property is relaxed to \emph{zero-egress-after-load}: model weights range from approximately 1\,GB (small quantized models) to 8\,GB+ (larger models), making true offline-first impractical for capable models. However, once weights are cached, the application functions without network access and with complete zero-egress guarantees.

This implementation also introduces WebGPU as a new execution substrate. While FFmpeg (WASM) and Whisper (ONNX via Transformers.js) run primarily on the CPU, WebLLM leverages the GPU for matrix multiplication via WebGPU compute shaders, achieving inference speeds that approach those of dedicated local LLM runtimes. This demonstrates that the Web-CLI pattern is not limited to CPU-bound workloads but extends to GPU-accelerated inference.

The privacy implications of client-side LLM inference are arguably the most profound of our case studies. Sending a prompt to a remote language model reveals not just file contents (as with video editing) or spoken words (as with transcription) but cognitive state: concerns, intentions, questions, and reasoning. A Web-CLI LLM is the only architecture in which this information never leaves the device by construction.

\subsection{3D Print Preparation: A Physical Output}
\label{sec:3mf}

Our fourth reference implementation, \texttt{3mf-webCLI}~\cite{tdmfwebcli}, extends the pattern in a direction the first three do not: its output is not a media file consumed on screen but a manufacturing description consumed by a physical 3D printer. It converts a GLB (binary glTF) model into a multi-material 3MF package, segmenting the model's colors into a printer-appropriate palette (e.g., four materials for a Bambu AMS, or two-to-five for a Prusa MMU) so that a single mesh can be printed in multiple filaments. The processing is entirely deterministic: a perceptual color-segmentation pipeline (chroma-weighted Lab $k$-means with multi-restart, edge-aware label smoothing, and Markov-random-field refinement over the face-adjacency graph) followed by 3MF authoring. Like the media tools, it runs wholly in the browser: the model is parsed, segmented, and packaged on the client, and neither the input model nor the output ever leaves the device.

This implementation is instructive for three reasons. First, it widens the pattern's domain from media and inference to \emph{geometry processing}, and its output modality from an on-screen file to a physical artifact, showing that the four defining properties are not specific to media or to neural workloads. Second, it is notably lighter than the media tools: it requires no large binary or model weights (only a 3D library loaded from a CDN and cached), no cross-origin isolation, and no special hosting headers, and so represents an especially pure instance of the thin-shell, client-compute economics described in Section~\ref{sec:cost}. Third, its privacy case is concrete and commercial: a proprietary or commissioned 3D model can be prepared for multi-material printing without being uploaded to a cloud converter or model library. It also composes with adjacent authoring tools: a scene edited in a browser-native 3D editor can be exported and prepared for printing entirely on-device, extending the on-device pipeline from capture and authoring through to physical output.

\subsection{Toward AI-Native Web-CLIs}
\label{sec:ainative}

The reference implementations above all follow the literal Web-CLI pattern: a graphical interface maps user actions to the arguments (or inference parameters, or processing configuration) of an underlying capability. The local-LLM case study, however, points toward a more general form of the architecture that we believe is the pattern's most significant future direction.

In a conventional Web-CLI, the interface translates user intent into tool arguments through hand-written mappings. An \emph{AI-native} Web-CLI replaces that mapping with a local language model: the user expresses intent in natural language, and the on-device model generates the commands or code that drive an underlying engine. The command line becomes a conversation, still conducted entirely on the user's hardware. Because the model runs locally, this preserves the zero-egress guarantee even as the interface becomes substantially more expressive. The user's natural-language intent, which may reveal far more than a file ever could, never leaves the device.

This generalization raises a fidelity challenge absent from deterministic Web-CLIs. A small, locally-runnable model is an unreliable command generator: it may invent non-existent API calls or produce arguments that fail. We anticipate that AI-native Web-CLIs will address this not by relying on larger models but with an architectural pattern: a bounded agentic loop of the form \emph{generate, validate, execute, observe, fix}. Generated commands are statically validated against a locally-indexed catalog of the real API surface (rejecting hallucinated calls before execution), executed through a single reversible command surface, and their effect on application state observed and fed back for correction. Two further on-device techniques make such a loop practical: retrieval of real API signatures from a local index injected before generation (which constrains the model to the genuine interface), and grounding the model in application state through the existing runtime (for example, querying a renderer directly for what is visible) rather than downloading an auxiliary model. Each of these mechanisms keeps the entire pipeline, including all verification, on the client, preserving zero egress while compensating for the limited capability of locally-runnable models. We regard this agentic, locally-verified interface as the natural endpoint of the progressive-disclosure principle: natural language at the top level, generated code in the middle, and a raw execution surface beneath, all on-device.

\subsection{Comparative Analysis}

Table~\ref{tab:casestudies} summarizes the four reference implementations against the Web-CLI properties and key technical characteristics.

\begin{table}[h]
\caption{Web-CLI case studies compared across key dimensions}
\label{tab:casestudies}
\small
\begin{tabularx}{\linewidth}{lXXXX}
\toprule
& \textbf{FFmpeg} & \textbf{Whisper} & \textbf{WebLLM} & \textbf{3mf-webCLI} \\
\midrule
Domain & Media processing & Speech recognition & LLM inference & 3D print prep \\
Execution & WASM (Emscripten) & ONNX/WASM (Transformers.js) & WebGPU runtime & JavaScript (deterministic) \\
Binary size & 31\,MB & 39\,MB--1.5\,GB & 1--8\,GB & 3D lib (CDN) \\
Fidelity & Full & Full & Full & Full \\
Prog. disclosure & Yes & Yes & Yes & Yes \\
Offline-first & Yes & Yes & Relaxed & Yes \\
Zero egress & Yes & Yes & Yes & Yes \\
Privacy sensitivity & High & Very high & Extreme & Commercial \\
\bottomrule
\end{tabularx}
\end{table}

%% ============================================================
\section{Evaluation}
\label{sec:evaluation}
%% ============================================================

\subsection{Performance Evaluation}

We benchmarked \texttt{ffmpeg-webCLI} against native FFmpeg across a duration sweep: 60-, 150-, and 600-second 1080p H.264 clips (24\,fps; 14.5, 33.8, and 56.9\,MB) created by stream copy, without re-encoding, from the same Big Buck Bunny source~\cite{bbb}\footnote{All 3 clips are available at \url{https://github.com/tejaswigowda/ffmpeg-webCLI/tree/main/benchmark/video}.}, so the input encoding is identical across lengths. Because a user's native baseline may be software or hardware-accelerated, we measured two native conditions: software x264 (\texttt{libx264}; the architectural apples-to-apples comparison, since the WASM build is software-only) and the platform hardware encoder (VideoToolbox) for operations that re-encode H.264. All conditions used identical FFmpeg arguments, namely those generated by the application's presets. Each cell reports the median of three runs (stream-copy operations and all native conditions) or two runs (WASM re-encodes; run-to-run spread below 1\%), with a discarded WASM warm-up execution preceding timed runs. The one exception is WASM GIF conversion at 150\,s, whose large intermediate palette buffers make it memory-sensitive: repeated runs in a single long-lived tab degrade as the heap fragments, so we report the median of the stable runs (which agree to within 1\%) from fresh page loads. Benchmarks ran on a MacBook Pro (M1 Max, 10 CPU cores, 64\,GB RAM) on AC power under macOS 26.6, comparing \texttt{ffmpeg-webCLI} in Chrome 150 (ffmpeg.wasm 0.12.15 with the single-threaded \texttt{@ffmpeg/core} 0.12.10, FFmpeg 5.1.4) against native FFmpeg 8.1.1 (Homebrew).

\begin{table}[h]
\caption{Median processing time: native FFmpeg (hardware and software encoders) vs. \texttt{ffmpeg-webCLI} (WASM). Ratios are WASM over the respective native condition.}
\label{tab:performance}
\begin{tabularx}{\linewidth}{lXrrrrr}
\toprule
\textbf{Operation} & \textbf{Clip} & \textbf{HW (s)} & \textbf{SW (s)} & \textbf{WASM (s)} & \textbf{vs SW} & \textbf{vs HW} \\
\midrule
MP4 to WebM (VP9) & 60\,s & -- & 110.7 & DNF\textsuperscript{\dag} & -- & -- \\
\midrule
Compress (H.264, CRF 28) & 60\,s & 4.0 & 13.1 & 325.0 & 24.9$\times$ & 81.0$\times$ \\
 & 150\,s & 9.6 & 22.8 & 690.2 & 30.2$\times$ & 72.2$\times$ \\
 & 600\,s & 37.3 & 81.5 & 2707.5 & 33.2$\times$ & 72.5$\times$ \\
\midrule
Rotate 90° (H.264) & 60\,s & 4.0 & 12.3 & 360.4 & 29.2$\times$ & 89.7$\times$ \\
 & 150\,s & 9.7 & 25.1 & 756.8 & 30.2$\times$ & 78.3$\times$ \\
 & 600\,s & 37.3 & 103.2 & 2976.8 & 28.8$\times$ & 79.7$\times$ \\
\midrule
Audio extraction (MP3) & 60\,s & -- & 0.44 & 0.58 & 1.3$\times$ & -- \\
\midrule
Convert to GIF (480\,px, 15\,fps) & 60\,s & -- & 28.2 & 40.6 & 1.4$\times$ & -- \\
 & 150\,s & -- & 69.2 & 108.0 & 1.6$\times$ & -- \\
 & 600\,s & -- & 261.9 & DNF\textsuperscript{\ddag} & -- & -- \\
\midrule
Trim (stream copy) & 60\,s & -- & 0.04 & 0.02 & 0.5$\times$ & -- \\
 & 150\,s & -- & 0.05 & 0.04 & 0.8$\times$ & -- \\
 & 600\,s & -- & 0.08 & 0.07 & 0.9$\times$ & -- \\
\midrule
Metadata strip (copy) & 60\,s & -- & 0.05 & 0.03 & 0.6$\times$ & -- \\
 & 150\,s & -- & 0.06 & 0.06 & 1.0$\times$ & -- \\
 & 600\,s & -- & 0.13 & 0.13 & 1.0$\times$ & -- \\
\bottomrule
\end{tabularx}

\smallskip
\footnotesize{\textsuperscript{\dag}Did not finish: VP9 encoding of 1080p input reproducibly exhausts the 2\,GB WASM heap. \textsuperscript{\ddag}Did not finish: GIF conversion of the 600\,s clip fails in the \texttt{palettegen} filter with an out-of-memory error (the filter buffers all $\sim$9{,}000 frames).}
\end{table}

Three regimes emerge. \textbf{Stream-copy operations} (trim, metadata strip) run at parity or faster in WASM ($0.5$--$1.0\times$): with an in-memory virtual filesystem, I/O-bound work carries no penalty. \textbf{Operations that are single-threaded natively} (MP3 extraction, GIF palette generation) run only modestly slower ($1.3$--$1.6\times$), consistent with prior measurements of WASM's per-instruction overhead~\cite{jangda2019}. \textbf{Natively multithreaded re-encodes} show the large gaps: H.264 encoding runs $25$--$33\times$ slower than software x264 and $72$--$90\times$ slower than VideoToolbox hardware encoding. The dominant factor is parallelism, not WASM itself: the deployed core is single-threaded while native x264 saturates all ten cores, so the implied per-core gap ($\approx$$2.5$--$3.3\times$) is consistent with the single-threaded measurements. The WASM-to-software ratio grows mildly with clip length ($24.9\times$ at 60\,s to $33.2\times$ at 600\,s for compression), and the absolute numbers make the practical envelope concrete: compressing a 10-minute 1080p clip takes 45 minutes in WASM versus 81 seconds natively. Two operations did not complete at the longest length: VP9 WebM conversion (at every length) and GIF conversion of the 600\,s clip, both reproducibly exhausting the 2\,GB WASM heap: VP9 in the encoder, GIF in the frame-buffering \texttt{palettegen} filter.

These results sharpen the applicability claim: for stream-copy, audio, and short-clip operations the Web-CLI is effectively free; for long re-encodes the current single-threaded deployment is one to two orders of magnitude slower than native. Closing that gap (a multi-threaded core, WebCodecs, or WebGPU encoding) is an engineering roadmap rather than an architectural limit.

\subsection{Total Time-to-Result}
\label{sec:totaltime}

Per-operation processing overhead, however, is not the metric a user experiences. The user-relevant quantity is \emph{total time-to-result}, and here the comparison is more subtle. For a cloud tool this total is the sum of four stages: upload, an unbounded and opaque server-side queue, processing, and download. For a Web-CLI it is a single stage: local processing. The file is already on the device and the result stays on it, so both transfer stages are eliminated entirely, and there is no queue.

The eliminated stages are often the dominant and least predictable costs. Upload and download times scale with file size and connection speed (minutes for a large clip on a typical residential uplink, worse on mobile), while the server-side queue is a black box: its wait depends on other users' load, is rarely reported to the user, may balloon precisely when a service is popular, and can fail after the user has already waited. Against these, a Web-CLI's higher-but-\emph{bounded} local processing time is knowable and fully under the user's control, with progress visible on their own machine. The measured sweep makes the envelope precise. For stream-copy, audio, and other at-parity operations (Table~\ref{tab:performance}), the Web-CLI wins outright: total time-to-result is essentially the operation itself, with zero transfer and zero queue. For short re-encodes (a 60-second compress: about 5.4 minutes locally), eliminating a multi-minute upload, an unbounded queue, and a download keeps the local path competitive on typical uplinks. For long re-encodes (a 10-minute compress: about 45 minutes locally), a fast cloud pipeline will often win end-to-end today; what the Web-CLI offers there is predictability, and for privacy-required users it is the only available path, since the cloud option, upload included, is simply unavailable to them. The re-encode gap is a property of the current single-threaded core rather than of the architecture (Section~\ref{sec:evaluation}).

\subsection{Feature Parity}

We evaluated feature parity against native FFmpeg by reproducing the 30+ preset operations and all 13 example command library recipes using native FFmpeg with identical arguments. Every tested operation produced output equivalent to the native implementation: identical for stream-copy operations, and perceptually equivalent for re-encoding operations (with minor bit-level differences expected due to threading and SIMD differences).

The one operation lacking a native equivalent is hard subtitle burning, due to the absence of \texttt{libass} in the standard \texttt{ffmpeg.wasm} build. As described in Section~\ref{sec:implementation}, the application provides this capability through a Canvas-based compositing workaround rather than FFmpeg's native subtitle filter; the workaround covers common burned-in captioning but not advanced \texttt{libass} features. This limitation is documented in the interface. The performance benchmark additionally surfaced two scale limitations that stem from the 2\,GB WASM heap rather than from missing features: VP9 WebM conversion of 1080p input does not complete at any tested length, and GIF conversion of the 600\,s clip exhausts memory in the \texttt{palettegen} filter (Table~\ref{tab:performance}); both commands complete natively. Feature parity at 1080p is therefore complete except for hard subtitle burning (worked around by design) and these two memory-bound cases.

\subsection{Accessibility and Progressive Disclosure}

We do not claim an empirical usability result; a controlled study of non-technical task completion is future work. We make instead a design argument. The barrier the Web-CLI removes is threefold: the tool need not be installed, no command-line syntax is required for common operations, and no account or upload is involved. Progressive disclosure (Section~\ref{sec:architecture}) is the mechanism: the graphical layer maps common tasks to constrained controls, so a user who has never opened a terminal can trim, convert, or caption a video, while the raw control surface beneath preserves full capability for experts. The same layered structure that makes the tool approachable for a novice keeps it uncompromised for a power user; and, as we note below, it exposes a machine-readable command surface that agents can drive as well.

\subsection{Independent Adoption}
\label{sec:adoption}

Beyond our own reference implementations, we have observed early signs of independent uptake: third-party browser tools that reproduce the pattern's architecture and framing (client-side execution via \texttt{ffmpeg.wasm}, no upload or account, files kept local), some of which attribute the interaction model to our work. We report this as anecdotal rather than systematic evidence; a rigorous study of downstream adoption is future work. Even so, external and unsolicited reuse is a useful signal that the pattern is reproducible and valued by developers who encountered it independently. We note that our own instantiations enforce the stricter, verifiable form of the zero-egress guarantee (no analytics, no server-backed fallback), whereas some adopters relax it to ``local where possible''; the pattern appears to spread even where its most rigorous privacy discipline is not fully retained.

\subsection{Privacy Verification}

We verified the zero-egress property by running the complete operation set under Chrome DevTools Network monitoring. Zero outbound requests were observed during any processing operation. The application makes network requests only on initial load (the \texttt{ffmpeg-core.wasm} binary and associated JavaScript) and for no other purpose.

%% ============================================================
\section{Discussion}
\label{sec:discussion}
%% ============================================================

\subsection{The Web Platform as a Rich Substrate}
\label{sec:platform}

A Web-CLI application is not merely a sandboxed binary; it is a web-native application, and as such it inherits the full and rapidly-growing capability surface of the modern web platform without installation, drivers, or platform-specific code. This is a substantive advantage over both traditional installed software and over the narrow ``WASM-in-a-sandbox'' view of browser computation.

Several browser APIs are directly relevant to Web-CLI applications:
\begin{itemize}
  \item \textbf{WebGPU} exposes the GPU for general-purpose compute, enabling on-device neural inference and accelerated media processing. Our speech-recognition and WebLLM case studies rely on it, and we confirmed GPU-accelerated inference not only on desktop but in mobile browsers on both Android and iOS -- transcribing a 56-minute lecture with the Whisper tiny model in 5 minutes on an M1 laptop and 13 minutes on an Android phone (the latter at roughly 1\% battery), entirely on-device. FFmpeg-class operations with GPU equivalents (encoding, scaling, color transforms) could be similarly accelerated, narrowing the gap with native execution.
  \item \textbf{Web Serial} and \textbf{Web Bluetooth (Web BLE)} grant direct access to local hardware (serial devices, microcontrollers, sensors, and BLE peripherals), entirely on-device. A Web-CLI tool can read from and control physical devices with no native driver and no intermediary server.
  \item \textbf{WebRTC} and \textbf{WebSockets} provide real-time peer-to-peer and full-duplex connectivity. Notably, WebRTC enables \emph{device-to-device} data exchange that need not traverse a central server, allowing collaborative or multi-device Web-CLI workflows that still avoid third-party data custody.
  \item \textbf{The File System Access API}, \textbf{Web Workers}, \textbf{IndexedDB}, and the \textbf{Web Audio API} round out a platform capable of handling large files, sustained background computation, persistent local caching, and real-time signal processing, all client-side.
  \item \textbf{3D Web and WebXR}. WebGL and WebGPU expose hardware-accelerated 3D rendering, and the WebXR Device API extends it to immersive augmented and virtual reality, all on-device. A Web-CLI can parse, render, and manipulate 3D assets (meshes, glTF/GLB scenes, point clouds) entirely in the browser, and preview results in AR before committing them. Our \texttt{3mf-webCLI} case study (Section~\ref{sec:3mf}) draws on exactly this substrate: it loads and segments 3D models locally, with no geometry ever leaving the device.
  \item \textbf{Multi-user connected 3D experiences}. Combining the 3D and real-time-connectivity primitives above, WebRTC and WebSockets data channels can synchronize scene state directly between participants, enabling collaborative 3D editing, shared spatial sessions, and multiplayer WebXR. Because peers can exchange state directly (Section~\ref{sec:platform}), such experiences remain multi-user without surrendering data custody to a central server: the shared payload is geometry and interaction state among chosen peers rather than an upload to a third party, so the collaboration is social while the zero-egress discipline toward external services is preserved.
\end{itemize}

This rich substrate has two consequences for the architecture. First, it expands the design space well beyond what a headless CLI tool traditionally offered: a Web-CLI can capture from a microphone, drive a serial device, accelerate on the GPU, or coordinate peer-to-peer, all while remaining a single zero-install web application. Second, it clarifies the precise nature of the zero-egress guarantee. Because the platform \emph{does} expose egress channels (\texttt{fetch}, WebSockets, WebRTC), zero egress is not a physical impossibility imposed by the sandbox but a verifiable design property of a well-constructed Web-CLI. The application simply does not transmit user data; unlike a remote service, this can be confirmed by direct inspection. The architecture's strength is therefore not that transmission is impossible, but that the absence of transmission is \emph{checkable} by anyone, at any time, in a way a server-side service can never offer.

A further consequence concerns reach. Because a Web-CLI is a Progressive Web App, it installs to the home screen and launches as a standalone application not only on desktop and Android but on iOS as well. This is significant precisely because iOS is the most tightly gatekept consumer platform: the App Store is normally the only sanctioned path to an installed application. A Web-CLI delivers an installed, GPU-accelerated, zero-egress application on iOS without the App Store, developer-account enrollment, or review process -- using only web standards. The web platform thus now reaches capabilities that previously required platform-gated native installation, even on the most restrictive devices. (We note one practical caveat: iOS may evict a PWA's cached storage under storage pressure, which can require re-downloading model or binary assets on next use; the zero-egress guarantee is unaffected.)

\subsection{Emergent Platform Capabilities}
\label{sec:emergent}

A defining design choice of the Web-CLI is to be \emph{web-native first}: the architecture is specified by its four properties and realized on the web platform, rather than by any fixed toolchain. A direct consequence is that Web-CLI applications inherit the capabilities of the web platform as \emph{emergent} benefits -- capabilities the application never implements itself, that arise simply from building on the platform, and that improve over time as the platform evolves.

Our reference implementations make this concrete. We did not write GPU kernels, yet transcription and LLM inference are GPU-accelerated through WebGPU. We did not write platform-specific installers, yet the applications install to the home screen and launch standalone on desktop, Android, and iOS through the Progressive Web App standard. We did not write hardware drivers, yet the platform exposes serial, Bluetooth, and filesystem access to any Web-CLI that needs them. Each of these is a capability a native application would have to implement, and on each platform separately; for a Web-CLI they emerge from the substrate.

Two aspects of this are worth emphasizing. First, the \emph{cross-platform} reach is arguably more distinctive than any single capability: one codebase yields a GPU-accelerated, installable application across desktop, Android, and iOS -- including, on iOS, an installed app delivered without the App Store. Platform fragmentation, the defining tax of native development, is avoided not through cross-platform frameworks but as a side effect of being web-native. Second, the architecture \emph{compounds with the platform}: GPU acceleration arrived with WebGPU and required no change to our applications; as standards such as the Web Neural Network API mature, on-device inference Web-CLIs stand to inherit dedicated NPU acceleration with equally little effort. The Web-CLI is therefore not a fixed point but a position on a rising tide -- its capabilities grow as the web platform does, without per-application engineering.

\subsection{Generalizability of the Web-CLI Pattern}

Our four case studies demonstrate the Web-CLI pattern across distinct computational domains, execution substrates (CPU/WASM and GPU/WebGPU), output modalities (on-screen media and physical manufacturing), and privacy threat models; early anecdotal signs of independent reuse (Section~\ref{sec:adoption}) suggest it applies beyond our own implementations. We argue that the pattern is broadly applicable to any tool satisfying the suitability criteria defined in Section~\ref{sec:architecture}.

Table~\ref{tab:candidates} lists additional tools we assess as strong Web-CLI candidates based on these criteria.

\begin{table}[h]
\caption{Candidate tools for future Web-CLI implementations}
\label{tab:candidates}
\begin{tabularx}{\linewidth}{lX}
\toprule
\textbf{Tool} & \textbf{Domain / Notes} \\
\midrule
ImageMagick & Image processing; large C codebase, WASM-compilable \\
Pandoc & Document conversion; Haskell, requires WASM Haskell runtime \\
Sox & Audio processing; C, straightforward WASM port \\
Ghostscript & PDF processing; C, active WASM experiments \\
SQLite & Database; official WASM build already ships \\
yt-dlp & Video download; Python, requires WASM Python runtime \\
\bottomrule
\end{tabularx}
\end{table}

SQLite deserves special mention as a prior instantiation of the Web-CLI pattern that predates our work. The SQLite team's official WASM deployment~\cite{sqlitewasm} satisfies all four Web-CLI properties and has seen wide adoption in browser-based applications. Our contribution is to abstract and name the pattern that SQLite (and \texttt{ffmpeg.wasm}) exemplify, and to demonstrate its applicability to a broader class of tools including neural inference workloads.

\subsection{The Web-CLI as Default Architecture for Sensitive Data}

A recurring theme across the case studies is that the most compelling use cases for Web-CLI applications involve data that users have strong reasons not to transmit to a third party. Medical video, legal recordings, journalistic sources, and personal communications are all cases where the zero-egress guarantee is not merely convenient but ethically and legally significant.

We argue that for this class of applications (those processing data whose disclosure to a third party would constitute a meaningful privacy violation), the Web-CLI should be considered the \emph{default} architecture. The burden of justification should fall on architects who choose to process this data server-side, not on those who choose client-side processing.

This is a stronger claim than is commonly made in the privacy-preserving computation literature, which tends to frame client-side processing as a premium feature available at the cost of performance or capability. The Web-CLI architecture demonstrates that for a broad class of operations (stream-copy edits, audio extraction, light transcoding, and short-clip work), client-side processing is available at little or no performance cost and with no capability reduction. Heavy re-encoding is the exception, running one to two orders of magnitude slower in the current single-threaded WASM build; but even there the traditional justification for server-side processing (that the client cannot bear the computational load) no longer holds categorically: the client bears it, more slowly, and for privacy-required data that is the only admissible option regardless of speed. The performance cost, where it exists, is a property of the current deployment rather than of the architecture.

\subsection{Cost and Distribution Structure}
\label{sec:cost}

A final property reinforces the case for the Web-CLI as a default: its cost and distribution structure. Because all computation occurs on the client, a Web-CLI has no per-user server cost. The deployment is a thin static shell (HTML, application JavaScript, and a service worker) served from free static hosting; the heavy assets (the WASM binary and model weights) are delivered by content-delivery networks and then cached on-device by the service worker, so each asset is fetched at most once per client. The marginal cost of an additional user is therefore effectively zero, and operational cost is largely independent of the number of users, in sharp contrast to server-bound tools whose processing cost scales directly with usage. In practice this is near-zero rather than literally zero (free static hosting carries soft bandwidth limits, and asset delivery consumes CDN quota), but it scales for free across the range that matters and can be sustained indefinitely without a revenue model. This is also the economic root of the absence of accounts, usage tiers, quotas, and advertising: with no server to sustain, there is nothing to authenticate against, meter, or monetize. Distribution follows the same logic. A Web-CLI is delivered by URL and installs as a Progressive Web App across desktop, Android, and iOS, reaching even the walled-garden platforms without app-store submission, review, or revenue share. The architecture thus supports sustainable, ungated, universal reach that a server-bound alternative cannot match.

\subsection{Limitations and Open Problems}

\textbf{Binary size and load time.} The 31\,MB initial load for \texttt{ffmpeg.wasm} is a meaningful barrier for users on slow connections. Streaming compilation (loading and compiling the WASM binary simultaneously) mitigates this, but does not eliminate it. Future work on WASM binary compression and modular loading could reduce this barrier.

\textbf{Memory constraints.} The 2\,GB WASM memory ceiling limits the tool to files that fit within this budget. For professional video workflows involving large files, this is a real limitation. Future browser implementations may raise or eliminate this ceiling.

\textbf{Supply chain trust.} The zero-egress guarantee applies to the deployed WASM binary. A malicious or compromised WASM binary could in principle exfiltrate data through browser APIs. Users who require a stronger trust model should build the WASM binary from source. Future work on formal verification of Web-CLI privacy properties is warranted.

\textbf{Codec licensing.} FFmpeg's inclusion of patented codecs (H.264, AAC) in its standard build creates licensing complexity for commercial deployments. This is a property of FFmpeg rather than the Web-CLI architecture, but it affects any Web-CLI implementation built on \texttt{ffmpeg.wasm}.

%% ============================================================
\section{Future Work}
\label{sec:futurework}
%% ============================================================

\subsection{GIMP as a Web-CLI: GUI-Native Tools}

Of the four implementations presented in this paper, three are headless tools whose interfaces are entirely command- or parameter-driven, and the fourth (\texttt{3mf-webCLI}) wraps a deterministic geometry pipeline; none begins as a GUI-native desktop application. An important open question is whether the Web-CLI pattern extends to tools that are fundamentally GUI-native.

GIMP represents the most compelling test case for this extension. As a professional image editor built on the GTK toolkit and the GEGL image processing library, GIMP presents challenges that FFmpeg does not: a full desktop GUI that must either be emulated or replaced, a plugin architecture spanning hundreds of independently developed extensions, and scripting runtimes (Script-Fu, Python-Fu) that themselves require porting.

We propose two approaches to a GIMP Web-CLI. The \emph{full-fidelity} approach compiles GIMP including its GTK interface to WASM and renders it via an HTML canvas backend, preserving the complete desktop experience. The \emph{progressive replacement} approach strips the GTK interface and replaces it with a web-native UI exposing GIMP's batch processing and Script-Fu scripting capabilities. This is analogous to our approach with FFmpeg. The privacy implications are significant: current browser-based image editors process images server-side. A GIMP Web-CLI would be the first privacy-preserving professional image editor deployable in the browser without installation.

\subsection{WebGPU Acceleration for Media Processing}

Current WASM-based media processing is CPU-bound. WebGPU enables GPU-accelerated compute for browser applications. FFmpeg operations with GPU equivalents (video encoding, scaling, color correction) could benefit substantially from WebGPU acceleration, narrowing the performance gap with native execution. A complementary path lies outside the browser: the same progressive-disclosure interface could wrap native FFmpeg in a desktop application with direct access to platform hardware encoders (e.g., VideoToolbox), collapsing the re-encode gap to native speed for the operations where the browser build is weakest. This trades the browser tier's two distinctive properties (zero-install distribution and in-browser verifiability) for hardware access, and so represents a different point on a capability-versus-reach spectrum rather than a replacement; the benchmark results (Section~\ref{sec:evaluation}) quantify precisely what such a native deployment would recover.

\subsection{A Web-CLI Toolkit}

The engineering effort required to port a CLI tool to the Web-CLI architecture is currently high and largely undocumented. A reusable toolkit (Emscripten build configurations, a standard JavaScript API layer, UI component library, and testing framework for verifying fidelity) would substantially lower this barrier. We intend to develop such a toolkit as a direct extension of this work.

\subsection{Formal Privacy Verification}

The zero-egress guarantee is currently argued informally. A formal treatment including a threat model accounting for supply chain attacks on the WASM binary, malicious CDN delivery, and browser extension interference would strengthen the privacy claim for adversarial settings.

\subsection{Broader CLI Tool Survey}

A systematic survey of widely-used CLI tools, assessing each against the four Web-CLI properties and identifying engineering barriers to porting, would provide the community with a roadmap for future implementations.

%% ============================================================
\section{Conclusion}
\label{sec:conclusion}
%% ============================================================

We have introduced the Web-CLI, a novel application architecture in which powerful computational capabilities (command-line tools compiled to WebAssembly, models run through client-side inference runtimes, and GPU-accelerated engines) are deployed as zero-install, offline-capable, privacy-preserving browser applications. We defined the architecture through four properties (fidelity, progressive disclosure, offline-first, and zero egress) and demonstrated its applicability across four reference implementations spanning deterministic media processing (FFmpeg), neural speech recognition (Whisper via Transformers.js), large language model inference (WebLLM), and geometry processing with a physical output (\texttt{3mf-webCLI}), with early anecdotal signs of independent reuse as suggestive further evidence that the pattern generalizes.

Our primary reference implementation, \texttt{ffmpeg-webCLI}, exposes 30+ FFmpeg operations through a web-native interface while preserving full access to FFmpeg's command-line interface. Evaluation shows output parity with native FFmpeg for the operation set, with a few disclosed exceptions (hard subtitle burning, worked around via Canvas compositing, and two memory-bound cases at 1080p: VP9 encoding at any length and GIF conversion of a 10-minute clip, blocked by the WASM heap ceiling). Performance ranges from native parity on stream-copy and audio operations to $25$--$33\times$ slower than multithreaded software encoding ($72$--$90\times$ versus hardware encoders) on H.264 re-encodes; the gap is dominated by the current core's single-threading rather than by WASM itself. The progressive-disclosure design lowers the barrier for non-technical users while preserving full capability for experts.

The Web-CLI architecture's most significant contribution is not performance or convenience but privacy. By making client-side execution the architectural default, it transforms the privacy guarantee from a policy commitment (which can be changed, violated, or subpoenaed) into a technical property that is independently verifiable and architecturally enforced. For the growing class of applications that process sensitive personal data, we argue this guarantee is not a luxury but a requirement, and that the Web-CLI architecture makes it achievable without sacrificing capability.

The browser is becoming a universal runtime. The Web-CLI is the pattern that makes it a privacy-preserving one.

%% ============================================================
%% Bibliography
%% ============================================================

\bibliographystyle{plainnat}
\bibliography{main}

\end{document}